\documentclass[aps,pra,showpacs,groupedaddress,floatfix,nofootinbib]{revtex4}
\usepackage{amsmath}
\usepackage{graphicx}
\usepackage{color}

\newcommand{\beq}{\begin{eqnarray}}
\newcommand{\eeq}{\end{eqnarray}}

\begin{document}

\title{{Analytical formulas for gravitational lensing}}
\author{Paolo Amore}
\email{paolo@ucol.mx}
\affiliation{Facultad de Ciencias, Universidad de Colima,\\
Bernal D\'{i}az del Castillo 340, Colima, Colima,\\
Mexico.}
\author{Santiago Arceo Diaz}
\affiliation{Facultad de Ciencias, Universidad de Colima,\\
Bernal D\'{i}az del Castillo 340, Colima, Colima,\\
Mexico.}

\begin{abstract}
In this paper we discuss a new method which can be used to obtain arbitrarily accurate 
analytical expressions for the deflection angle of light propagating in a given metric. 
Our method works by mapping the integral into a rapidly convergent series and provides 
extremely accurate approximations already to first order. We have derived a general 
first order formula for a generic spherically symmetric static metric tensor and we have tested it in four different 
cases. 
\end{abstract}

\pacs{98.62.Sb, 04.40.-b, 04.70.Bw}

\maketitle

\section{Introduction}

According to general relativity the trajectory of a ray of light which passes close 
to a mass distribution departs from being a straight line. The amount of the deflection
of the light depends upon the mass and can be quite large for light passing very close to 
a massive compact body, such as a black hole. 
The study of gravitational lensing under such conditions, also known as "strong gravitational 
lensing", has received wide attention in the recent past: for example strong gravitational 
lensing in a Schwarzschid black hole has been considered by Frittelli, Kling and 
Newman~\cite{FKN00} and by Virbhadra and Ellis~\cite{VE00}; Virbhadra and Ellis~\cite{VE02}
have later treated the strong gravitational lensing by naked singularities; Eiroa, Romero and
Torres\cite{Eir02} have described Reissner-Nordstr\"om black hole lensing, while Bhadra has considered
the gravitational lensing due to the GMGHS charged black hole~\cite{Bha03}; Bozza has studied
the quasiequatorial gravitational lensing by a spinning black hole~\cite{Boz03}; Whisker~\cite{Whi05} and
Eiroa~\cite{Eir05} have considered strong gravitational lensing by a braneworld black hole; 
still Eiroa~\cite{Eir06} has recently considered the gravitational lensing by an Einstein-Born-Infeld 
black hole; Sarkar and Bhadra have studied the strong gravitational lensing in the
Brans-Dicke theory\cite{SB06}; finally Perlick~\cite{Perl04} has obtained an exact gravitational lens equation
in a spherically symmetric and static spacetime and used to study lensing by a Barriola-Vilenkin monopole and
by an Ellis wormhole.

Different strategies have been used to evaluate the effects of strong gravitational lensing:
for example, Bozza~\cite{Boz02} has introduced an analytical method which allows to discriminate among 
different types of black holes: the method is based on a careful description of the logarithmic 
divergence of the deflection angle (the photon sphere); Mutka and M\"ah\"onen~\cite{Mutkaa,Mutkab}
and Belorobodov~\cite{Belo02} have derived improved formulas for the deflection angle in a Schwarzschild metric;
more recently, Keeton and Petters\cite{Keet05} have also developed a formalism for computing corrections 
to lensing observables in a static and spherically symmetric metric beyond the weak deflection limit.

The purpose of this paper is to present a new method which can be used to calculate analytically and
systematically the deflection angle in a static spherically symmetric metric. Originally this method
was devised by Amore and collaborators~\cite{Am05a,Am05b} to obtain analytical formulas for the 
period of a classical oscillators: our method works by converting the integral which needs to be calculated into
a series depending upon a variational parameter. Such procedure is inspired by the Linear Delta 
Expansion (LDE) method~\cite{lde} and by Variational Perturbation Theory~\cite{VPT}. 
For certain values of the variational parameter, the series obtained is proved to converge to the 
exact result, while at finite orders a particular value of the parameter can be chosen using the 
Principle of Minimal Sensitivity (PMS)~\cite{Ste81} to minimize the error. Fully analytical results,
which do not correspond to a perturbative expansion in some small parameter, are obtained.

The paper is organized as follows: in Section \ref{method} we describe the method and obtain a general first
order formula, which is valid for an arbitrary metric tensor; in Section \ref{conv} we discuss the convergence
of the method and provide an estimate of the rate of convergence, which is proved to be exponential;
in Section \ref{appl} we apply our formula  to four different metric tensors and discuss the precision of our approximation, comparing it with
the available results in the literature; finally in Section \ref{concl} we draw our conclusions.

\section{The method}
\label{method}

We are interested in the general static and spherically symmetric metric which corresponds to
the line element
\beq
ds^2 = B(r) dt^2 - A(r) dr^2 - D(r) r^2 \left(d\theta^2+ \sin^2\theta\ d\phi^2\right) 
\label{eq_1_1}
\eeq
and which contains the Schwarzschild metric as a special case. We also assume that the flat 
spacetime is recovered at infinity, i.e. that $\lim_{r\rightarrow \infty} f(r)  = 1$, where 
$f(r) = (A(r),B(r),D(r))$.

The angle of deflection of light propagating in this metric can be expressed by means of the integral~\cite{Weinberg}
\beq
\Delta\phi = 2 \int_{r_0}^\infty \sqrt{A(r)/D(r)} 
\sqrt{ \left[\left( \frac{r}{r_0}\right)^2 \frac{D(r)}{D(r_0)} 
\frac{B(r_0)}{B(r)}-1\right]^{-1}} \frac{dr}{r} - \pi \ ,
\label{eq_1_2}
\eeq
where $r_0$ is the distance of closest approach of the light to the center of the gravitational attraction.  

If we perform a change of variable, $z=r_0/r$, and define the function
\beq
V(z) &\equiv& z^2 \frac{D(r_0/z)}{A(r_0/z)} - \frac{D^2(r_0/z) \ B(r_0)}{A(r_0/z) B(r_0/z) D(r_0)} + 
\frac{B(r_0)}{D(r_0)} 
\eeq
we can write eq.~(\ref{eq_1_2}) in the form
\beq
\Delta\phi = 2 \int_0^{1} \frac{dz}{\sqrt{V(1)-V(z) }}-\pi \ .
\label{eq_1_5}
\eeq

Notice that $t = \sqrt{2} \int_0^1 dz/\sqrt{V(1)-V(z)}$ is the time spent by a classical oscillator 
moving in a potential $V(z)$ for passing from $z=0$ to the inversion point located at $z=1$; 
in a flat spacetime, where $A(r)=B(r)=D(r)=1$, $V(z)$ reduces to the familiar harmonic oscillator 
potential and the deflection angle identically vanishes. The integral of eq.~(\ref{eq_1_5}) 
can also be performed exactly in the case of the Schwarzschild metric~\cite{Darw59} and of the 
Reissner-Nordstr\"om metric~\cite{Eir02}. In both cases the exact result is expressed in terms of 
elliptic integrals. However, in more general cases the integration of eq.~(\ref{eq_1_5}) cannot be done 
exactly and one typically resorts to an expansion around the flat metric (of course such expansion can also
be used in cases where exact results are available, to avoid dealing with complicated special functions).
In the case of the Schwarzschild metric, for example, this approach yields a perturbative series in 
powers of $GM/r_0$, whose leading term $\Delta\phi_0 = 4 GM/r_0$ was first obtained by Einstein.
A big disadvantage of this approach is that the validity of the expressions obtained in this way
is restricted to large distance/weak field regime: for example, the exact solution to lensing in the 
Schwarzschild metric provided by Darwin possesses a singularity at $r_0 = 3 GM$, known as photon sphere,
and is clearly out of reach in a perturbative approach.

We will therefore pursue a different approach to deal with eq.~(\ref{eq_1_5}), which is not based 
on a perturbative expansion and which we will prove to be capable to describe very accurately the 
physics of our problem. The method that we propose has been devised by Amore and 
collaborators\cite{Am05a,Am05b} to obtain precise analytical formulas for the period of classical 
oscillators. In a recent work, the method has also been used to obtain analytical expressions for the
spectrum of quark-antiquark potentials~\cite{ADL06}.
A similar technique has also been applied by Amore to accelerate the convergence of certain
series (such as the Riemann and Epstein zeta functions) \cite{AmJMAA}, which can be used in the calculation of
loop integrals in finite temperature problems occurring in field theory~\cite{AmJPA}.

We now briefly describe our method. In the spirit of the Linear Delta Expansion we 
interpolate the full potential $V(z)$ with a solvable potential $V_0(z)$\footnote{By solvable
we mean that integrals $\int_0^1 z^n dz/\sqrt{V(1)-V(z)}$ can be performed analytically.}:
\beq
V_\delta(z) \equiv V_0(z) + \delta (V(z)-V_0(z)) \nonumber \ .
\eeq

Depending on the value of $\delta$ one will obtain the original potential ($\delta=1$) 
or the solvable potential ($\delta=0$).
In general $V_0(z)$ will depend upon one or more arbitrary parameters, which we will 
call $\lambda$: in fact we will assume the form $V_0(z) = \lambda z^2$.
With this definition we write the deflection angle as
\beq
\Delta\phi_\delta = 2 \int_0^{1} \frac{dz}{\sqrt{V_0(1) -V_0(z) + \delta ( V(1)-V(z)-V_0(1)+V_0(z)) }}-\pi \ ,
\label{eq_1_6}
\eeq
which clearly reduces to the original expression for $\delta=1$.

After introducing 
\beq
\Delta(z) \equiv \frac{E-V(z)}{E_0-V_0(z)}-1 \ .
\eeq
one can write  (\ref{eq_1_6}) as
\beq
\Delta\phi_\delta = 2 \int_0^{1} \frac{dz}{\sqrt{E_0 -V_0(z)}} \frac{1}{\sqrt{1 + \delta \Delta(z) }}-\pi \ .
\label{eq_1_7}
\eeq

Provided that $|\Delta(z)|<1$ for $0\leq z \leq1$ one can expand eq.~(\ref{eq_1_7}) in powers of 
$\delta$ and obtain a series (after performing the integrals) which converges to the exact result.
As discussed in \cite{Am05a,Am05b} this condition  requires that  $\lambda$ be greater than a 
critical value, $\lambda>\lambda_C$: in this case one obtains a family of series which depend 
upon $\lambda$ and which all converge to the exact result, which is independent of $\lambda$. 
However, if the series is truncated to a finite order, the partial sum displays an artificial 
dependence on $\lambda$: such dependence can be minimized by applying the Principle of Minimal 
Sensitivity (PMS)\cite{Ste81}: 
\beq
\frac{\partial}{\partial \lambda} \Delta\phi^{(N)} = 0 \ ,
\label{eq_1_8}
\eeq
having called $\Delta\phi^{(N)}$ the partial sum to order $N$. 

Notice that the solution to this equation selects the value of $\lambda$ where the series is 
less sensitive to changes in $\lambda$ itself: this value of $\lambda$ selects the series with 
the optimal convergence. In \cite{Am05a,Am05b} a large class of oscillators was studied 
using this method and it was found that our PMS series has an exponential rate of convergence.

However, since it was observed that the first order results are quite precise, we focus our present
effort in obtaining a first order formula, which is valid for a generic spherically symmetric static 
metric tensor corresponding to a potential 
\beq
V(z) = \sum_{n=1}^\infty v_n z^n \ .
\eeq

After expanding to first order we obtain 
\beq
\Delta\phi^{(1)} = \frac{2}{\sqrt{\lambda}} \int_0^{1} \frac{dz}{\sqrt{1-z^2}} 
\left[ 1 - \frac{ \Delta(z)}{2} \right] -\pi \ .
\label{eq_1_9}
\eeq
where
\beq
\Delta(z) =  
\sum_{n=1}^\infty \frac{v_n}{\lambda} \ \sum_{k=0}^{n-1}  \frac{z^k}{1+z} -1 \ .
\label{eq_1_10}
\eeq

The deflection angle can now be written as
\beq
\Delta\phi^{(1)} = \frac{3\pi}{2\sqrt{\lambda}} - \frac{1}{\lambda^{3/2}} 
\sum_{n=1}^\infty v_n \ \sum_{k=0}^{n-1}  I_k -\pi \ ,
\label{eq_1_11}
\eeq
where we have defined
\beq
I_{k} \equiv \int_0^{\pi/2} \frac{\sin^k\theta}{1+\sin\theta}  d\theta = \frac{\sqrt{\pi}}{\Gamma((k+1)/2)} \ 
\left[ \Gamma(k/2+1) - \frac{\left[\Gamma((k+1)/2)\right]^2}{\Gamma(k/2)} \right] \ .
\label{eq_1_12}
\eeq

Notice that eq.~(\ref{eq_1_11}) can be expressed in terms of the "transformed" potential
\beq
\rho(z) \equiv \sum_{n=1}^\infty v_n \Omega_n z^n \ ,
\label{eq_1_14}
\eeq
where
\beq
\Omega_n \equiv \sum_{k=0}^{n-1} I_k = \sqrt{\pi} \frac{\Gamma(n/2+1/2)}{\Gamma(n/2)} \ .
\label{eq_1_13}
\eeq

As expected our first order result depends upon $\lambda$ and we must use the PMS to obtain
the optimal value of $\lambda$:
\beq
\lambda_{PMS}^{(1)} = \frac{2\rho(1)}{\pi}  \ ,
\label{eq_1_16}
\eeq

Once this value is substituted inside eq.~(\ref{eq_1_11}) we obtain our first order 
result 
\beq
\Delta\phi_{PMS}^{(1)} = \pi \left[ \sqrt{\frac{\pi}{2\rho(1)}}- 1 \right] \ ,
\label{eq_1_17}
\eeq
which in our opinion is the most valuable formula contained in this paper. 
Notice that because of the form of eq.~(\ref{eq_1_17}) our approximation does not correspond 
to a perturbative expansion in some small parameter, as it will be clear in the next section.

\section{Convergence}
\label{conv}

In this section we discuss the convergence of our method: we wish to prove that our procedure 
provides series with an exponential rate of convergence.
As we have mentioned in the previous section we can write the expression for the deflection angle 
in a power series in $\Delta(z)$ as
\beq
\Delta \phi = 2 \int_0^1 \frac{dz}{\sqrt{E_0-V_0(z)}} \ \sum_{n=0}^\infty \frac{(2 n-1)!!}{2^n n!} \Delta^n(z) - \pi
\equiv \sum_{n=0}^\infty \Delta_n \ ,
\eeq
if the condition $|\Delta(z)|<1$ is fulfilled in the region of integration, $z \in (0,1)$.

Let us now call $\Delta_{max}$ the maximum value of $|\Delta(z)|$ in the region of integration; we can 
therefore write
\beq
|\Delta_n| \leq  \pi \frac{(2 n-1)!!}{2^n n!} \ \Delta_{max}^n \ .
\eeq

For large values of $n$ one can approximate the factorial and double factorial in this expression with the
corresponding asymptotic series, $(2 n-1)!!/2^n n! \approx 1/\sqrt{\pi n}$ and therefore
obtain
\beq
|\Delta_n| \leq  \frac{\Delta_{max}^n}{\sqrt{\pi n}}  \ ,  
\eeq
which confirms that the series converges geometrically. Notice that since $\Delta_{max}$ is defined 
in terms of the original potential $V(z)$ and of the interpolating potential $V_0(z)$, it will depend 
upon the arbitrary parameter $\lambda$: we therefore expect that the optimal value of $\lambda$, obtained 
at a finite order using the PMS, will be such that at large orders $\Delta_{max}$ assumes the smallest
value possible. Alternatively, one could think of lowering the value of $\Delta_{max}$ by choosing a 
potential different from the simple harmonic potential, $V_0(z) =  \lambda z^2$, discussed in the previous
section: indeed the only limitation that we have provided over $V_0(z)$ is that it is such that the integrals
contained in the series for $\Delta\phi$ can be performed analytically. This strategy, although possible, 
is not followed here because the increased complication in the form of $V_0(z)$ would necessarily reflect 
in a complication of the formulas obtained and in the drawback of obtaining approximations in terms of 
special functions. Clearly, such a procedure should also be investigated in future works.

We will now provide an estimate of $\Delta_{max}$:  under the assumption that $\Delta(z)$ is a monotonous 
function, we have that
\beq
\Delta_{max} = {\rm max} \left\{ |\Delta(0)|,|\Delta(1)|\right\} \ .
\eeq

The convergence of the series requires that $\Delta_{max} < 1$, which is fulfilled for
\beq
\lambda > \lambda_C = {\rm max} \left\{ \frac{1}{4} \sum_{n=1}^\infty n v_n  , \frac{1}{2} \sum_{n=1}^\infty v_n  \right\}\ .
\eeq

On the other hand it is easy to convince oneself that the minimal value of $\Delta_{max}$ is obtained when
the condition $\Delta(0) = - \Delta(1)$ is fulfilled, i.e. when 
\beq
\lambda_{max} = \frac{1}{2} \ \sum_{n=1}^\infty v_n  \left( 1 + \frac{n}{2}\right) \ ,
\eeq
corresponding to 
\beq
\Delta_{max} = \frac{\sum_{n=1}^\infty v_n}{\sum_{n=1}^\infty \frac{v_n}{2} (1+n/2)}-1 \ .
\eeq

The reader should also notice that our series cannot be used when $\lambda_{max} \leq \lambda_C$, because the condition
$\Delta_{max} < 1$ cannot be obeyed.

We can easily test our results over the Duffing potential $V(z) = z^2 +z^4$; in this case we have that $\lambda_{max}$ 
and $\lambda_{PMS}$ coincide ($\lambda_{PMS} = 5/2$). Corresponding to this value we have $\Delta_{max} = 1/5$ 
and we obtain a rate of convergence which is stronger than $r_n \approx (1/5)^{n}/\sqrt{\pi n}$, 
is in good agreement with the rate observed fitting the behavior of the series up to order $10$, 
$r_n \approx 0.028 \ \times \ 0.17^n$.

\section{Applications}
\label{appl}

We consider in this section four applications of the formula (\ref{eq_1_17}) obtained in 
the previous section. In the first two cases explicit formulas for the exact results are known
due to Darwin\cite{Darw59} and Eiroa and collaborators\cite{Eir02}; in the last two cases we consider
the metric of Janis-Newman-Winicour and the metric of a charged black hole coupled to Born-Infeld 
electrodynamics, for which no explicit formula is available.

\subsection{Schwarzschild metric}
\label{sbh}

Our first application is to the Schwarzschild metric, which corresponds to
\beq
B(r) = A^{-1}(r) = \left(1-\frac{2GM}{r}\right)  \ \ , \ \ D(r) = 1 \ .
\label{eq_2_1}
\eeq
Here $M$ is the Schwarzschild mass. The angle of deflection of a ray of light reaching a minimal distance
$r_0$ from the black hole can be obtained using eq.~(\ref{eq_1_5}).
The exact result can be expressed in terms of incomplete elliptic integrals of the first kind\cite{Darw59}
and reads
\beq
\Delta\phi = 4 \sqrt{\frac{\overline{r}_0}{\Upsilon}} \ \left[ F\left(\frac{\pi}{2}, \kappa\right) - 
F\left(\varphi, \kappa\right)  \right] \ ,
\label{eq_2_3}
\eeq
where $\overline{r}_0\equiv r0/GM$ and
\beq
\Upsilon \equiv \sqrt{\frac{\overline{r}_0-2}{\overline{r}_0+6}} \ \ &,& \ \  
\kappa \equiv \sqrt{(\Upsilon-\overline{r}_0+6)/2\Upsilon} \ \ , \ \ 
\varphi \equiv \sqrt{\arcsin \left[\frac{2+\Upsilon-\overline{r}_0}{6+\Upsilon-\overline{r}_0}\right]} \ .
\label{eq_2_4}
\eeq

Although eq.~(\ref{eq_2_3}) is exact, it is often valuable to obtain approximations which do not 
involve special functions. Here we will compare our first order approximation, corresponding to 
using eq.~(\ref{eq_1_17}), with other approximations which have been derived in the literature.

For example, Mutka and M\"ah\"onen~\cite{Mutkaa,Mutkab} have obtained the approximate formula
\beq
\Delta \phi_{MM} = \frac{4}{b-3} \ ,
\label{eq_2_5}
\eeq
where $b = r_0 \sqrt{D(r_0)/B(r_0)} = r_0/\sqrt{1-2 GM/r_0}$ is the impact parameter. 
This formula is a natural extension of the Einstein formula
\beq
\Delta \phi_{E} = \frac{4}{b} \ .
\label{eq_2_6}
\eeq
Beloborodov\cite{Belo02} has obtained another approximate formula which reads
\beq
\Delta \phi_{B} = \frac{4 GM}{r_0-2 GM} \ .
\label{eq_2_8}
\eeq

Finally, Keeton and Petters\cite{Keet05} have devised a systematic approach to deal with 
integrals as (\ref{eq_1_5}) and obtained the formula
\beq  
\Delta\phi_{KP} &=& A_1 \left(\frac{GM}{b}\right) + A_2 \left(\frac{GM}{b}\right)^2 + 
A_3 \left(\frac{GM}{b}\right)^3 + A_4 \left(\frac{GM}{b}\right)^4 \nonumber \\
&+& A_5 \left(\frac{GM}{b}\right)^5 +  A_6 \left(\frac{GM}{b}\right)^6 +  
O \left[ \left(\frac{GM}{b}\right)^7 \right] \ ,
\eeq
where the numerical values of the coefficients $A_i$ are given in eq.(25) of \cite{Keet05}.

Using the general equation for the deflection angle to first order, eq.~(\ref{eq_1_17}), 
we have obtained the formula:
\beq
\Delta \phi_{PMS}^{(1)} = \pi \ \left( \frac{1}{\sqrt{1-8 GM/\pi r_0}}-1 \right) ,
\label{eq_2_7} 
\eeq
corresponding to $\lambda_{PMS}^{(1)} = \sqrt{1-8 GM/\pi r_0}$.

\begin{figure}
\begin{center}
\includegraphics[width=9cm]{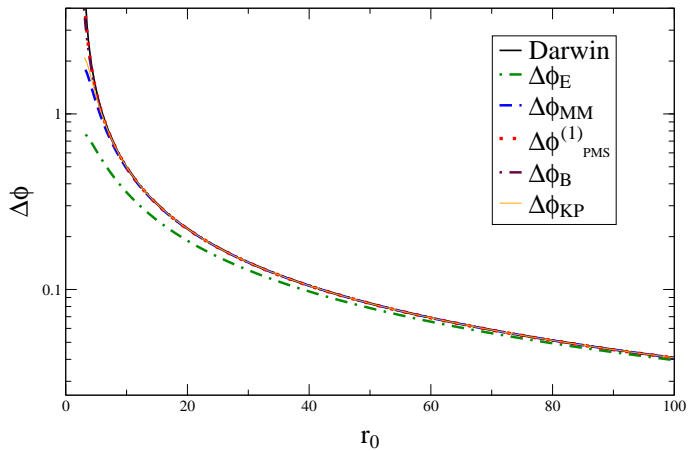}
\bigskip
\caption{Deflection angle as a function of $r_0$ assuming $GM=1$. (color online)}
\bigskip
\label{FIG1}
\end{center}
\end{figure}

\begin{figure}
\begin{center}
\includegraphics[width=9cm]{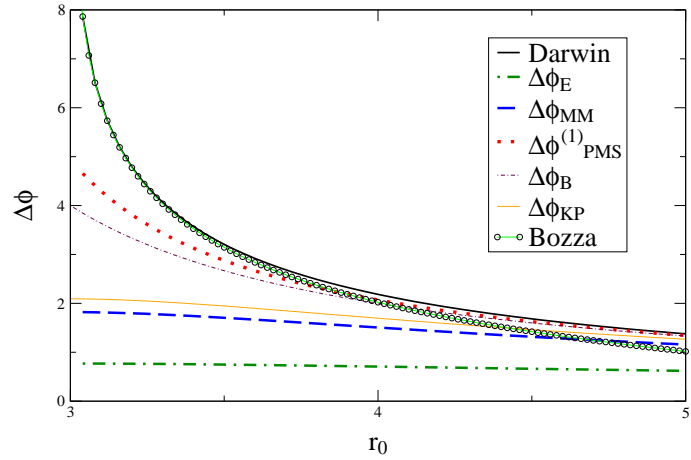}
\bigskip
\caption{Deflection angle as a function of $r_0$ assuming $GM=1$ close to the photon sphere. (color online)}
\bigskip
\label{FIG1b}
\end{center}
\end{figure}

Despite its simplicity, we can appreciate from Fig.~\ref{FIG1} and ~\ref{FIG1b} that 
eq.~(\ref{eq_2_7}) provides the best approximation to the deflection angle, even in 
proximity of the photon sphere (the singularity): indeed our formula predicts the location 
of the singularity at  $r_s = 8 GM/\pi \approx 2.55 GM$, slightly below the exact value 
$r_s^{(ex)} = 3 GM$. While the expression of Beloborodov puts the singularity at a smaller 
value of $r_0$, the remaining approximations either put it in the unphysical region 
($r_0 < 0$) ( Mutka and M\"ah\"onen) or fail to produce a singularity (Keeton and Petters).

In Fig.~\ref{FIG1b} we have also plotted the analytical approximation of Bozza~\cite{Boz02}, which 
correctly describes the photon sphere: our first order formula provides better approximations
already for $r_0 > 4 GM$.

Remarkably our expression works very well also in the opposite regime, corresponding to 
$r_0 \rightarrow \infty$; our eq.~(\ref{eq_2_7}) can be expanded for $r_0 \gg 1$ to give
\beq
\Delta \phi_{PMS}^{(1)} \approx \frac{4 GM}{r_0} + \frac{7.63944 G^2 M^2}{r_0^2} + \frac{16.2114 G^3 M^3}{r_0^3} + 
O\left[\left(GM/r_0\right)^4\right] \ ,
\eeq
which compares quite favorably with the exact asymptotic behaviour of the Darwin solution:
\beq
\Delta \phi \approx \frac{4 GM}{r_0} + \frac{7.78097 G^2 M^2}{r_0^2} + \frac{17.1047 G^3 M^3}{r_0^3} + 
O\left[\left(GM/r_0\right)^4\right] \ .
\eeq

\begin{figure}
\begin{center}
\includegraphics[width=9cm]{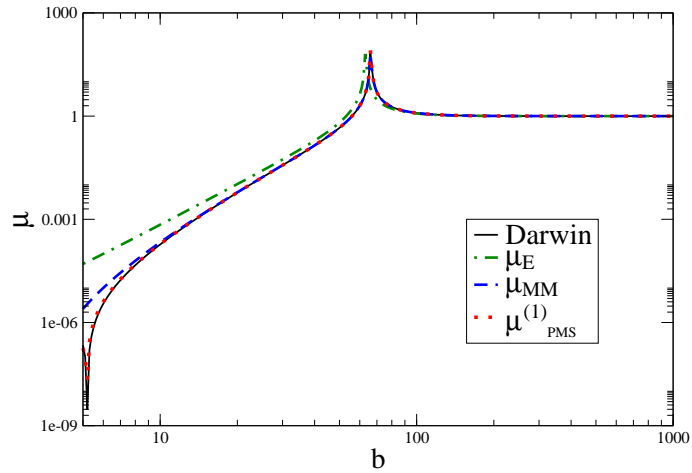}
\bigskip
\caption{Magnification as a function of the impact parameter assuming $a=1000$ and $GM=1$. (color online)}
\label{FIG2}
\bigskip
\end{center}
\end{figure}

In Fig.~\ref{FIG2} we also show the magnification (see \cite{Mutkaa})
\beq
\mu = \left| \left(1+ \frac{a}{b} \Delta\phi \right) \left( 1 + a \frac{\partial r_0}{\partial b} \frac{\partial \Delta\phi}{\partial r_0} \right)
\right|^{-1} \ ,
\eeq
as a function of the impact parameter $b$. $a$ is the distance between the lens and the source. 
Once again our simple formula provides a very accurate approximation to the exact result over a 
wide range of values.

Given the success of our approach to first order, a higher order calculation is not essential, 
although it is not technically difficult\footnote{In \cite{Am05a}, for example, the method was 
applied up to order $100$ to calculate analytically the period of an anharmonic oscillator 
with a precision of about $10^{-50}$.}. 
For example, it is straightforward to obtain the second order formula
\beq
\Delta \phi_{PMS}^{(2)} = \frac{\pi  \left(15 \pi^2 \left(\Lambda^2-1\right)^2-16 \pi  \left(\Lambda^2-1\right)^2-
32 \left(8 \Lambda^5-5 \Lambda^4-6 \Lambda^2+3\right)\right)}{256 \Lambda^5}
\eeq
where 
\beq
\Lambda &\equiv& \sqrt{\lambda_{PMS}^{(1)}} = \sqrt{1 - \frac{8 G M}{\pi r_0}} \ .
\eeq
This formula approximates the deflection angle to a $1 \%$ level up to $r_0 \approx 3.5 \ GM$, i.e.
quite close to the singularity, compared to $r_0 \approx 7.4 \ GM$ of the first order result.

Using the results obtained in Section \ref{conv} we can also estimate the rate of convergence of our series. In this case
the potential is 
\beq
V(z) = z^2 - \frac{2 GM}{r_0} z^3 \ ,
\eeq
and 
\beq
\Delta_{max} = \frac{GM/r_0}{2-5 GM/r_0} \ .
\eeq
It is interesting to notice that the condition of applicability of our series, $\lambda_{max} > \lambda_C$, can
be fulfilled only for $r_0 > 3 GM$, which is the exact location of the photon sphere for the Schwarzschild
metric: in other words, our series can also describe strong gravitational lensing close to the photon sphere.

\subsection{Reissner-Nordstr\"om metric}

The Reissner-Nordstr\"om (RN) metric describes a black hole with charge and corresponds to 
\beq
B(r) = A^{-1}(r) =  \left(1-\frac{2 GM}{r}+\frac{Q^2}{r^2}\right) \  \  &,& \ \ D(r) = 1 \ .
\eeq

As for the Schwarzschild metric the angle of deflection of a ray of light reaching a 
minimal distance $r_0$ from the black hole can be obtained from eq.~(\ref{eq_1_2}). 
Eiroa, Romero and Torres \cite{Eir02} have been able to express the deflection angle in 
terms of elliptic integrals of the first kind (see eqn. (A3) of  \cite{Eir02}).

It is straightforward to use our general formula to obtain the transformed potential for the 
RN metric
\beq
\rho(1)= \frac{\pi}{2} - \frac{4 GM}{r_0} + \frac{3 \pi Q^2}{4 r_0^2} 
\eeq
and thus the deflection angle
\beq
\Delta \phi_{PMS}^{(1)} = \pi \left[ \frac{1}{ \sqrt{ 1 - \frac{8 GM}{\pi r_0} + 
\frac{3  Q^2}{2 r_0^2} }}- 1 \right] \ .
\eeq

\begin{figure}
\begin{center}
\includegraphics[width=9cm]{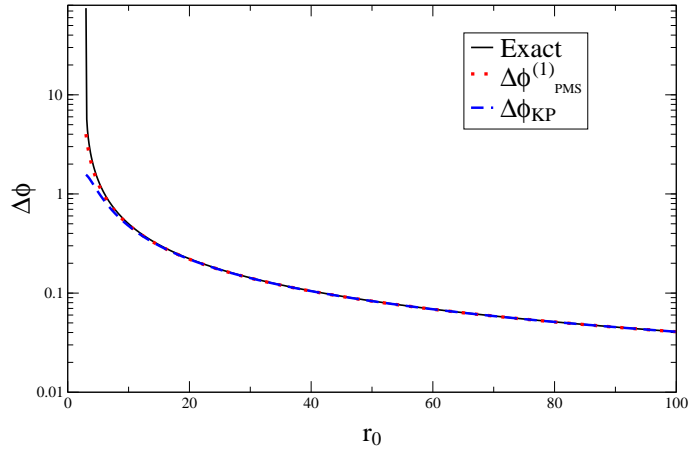}
\bigskip
\caption{Deflection angle for the Reissner-Nordstr\"om metric as a function of $r_0$ assuming $GM=1$ and 
$Q=1/2$. (color online)}
\label{FIG3}
\bigskip
\end{center}
\end{figure}

We can compare our formula both with the exact analytical result of Eiroa et al. and with the expressions 
(47) and (53) of \cite{Keet05}, which provide a systematic expansion of the deflection angle in 
terms of $GM/b$. In Fig.~\ref{FIG3} we have plotted the exact solution of \cite{Eir02} together 
with our first order formula and with the expression of Keeton and Petters, assuming $G=M=1$ and 
$Q=1/2$\footnote{Notice the different definition of $Q$ in \cite{Keet05}.}: the reader can appreciate that 
our simple formula is very accurate even in proximity of the photon sphere.

Notice that our expression reproduces well also the asymptotic behaviour of the deflection angle. 
In fact, eq.~(19) of \cite{Bha03} provides the leading asymptotic behavior of $\Delta\phi$, valid 
for $r \rightarrow \infty$:
\beq
\Delta\phi &\approx& \frac{4 GM}{r_0} + \frac{4 G^2M^2}{r_0^2} \left(\frac{15\pi}{16}-1\right) - 
\frac{3\pi}{4} \frac{Q^2}{r_0^2} + O\left[\left(\frac{1}{r_0}\right)^2\right]  \nonumber \\
&\approx& \frac{4 GM}{r_0} + 7.78 \ \frac{G^2M^2}{r_0^2} - 2.36 \ \frac{Q^2}{r_0^2} + 
O\left[\left(\frac{1}{r_0}\right)^2\right] \  .
\eeq
which can be compared with the asymptotic behaviour of our formula
\beq
\Delta\phi_{PMS}^{(1)} &\approx&  \frac{4 GM}{r_0} + \frac{24 G^2M^2}{\pi r_0^2} - \frac{3\pi}{4} \frac{Q^2}{r_0^2} +
O\left[\left(\frac{1}{r_0}\right)^2\right]  \nonumber \\
&\approx& \frac{4 GM}{r_0} + 7.64 \ \frac{G^2M^2}{r_0^2} - 2.36 \ \frac{Q^2}{r_0^2} + 
O\left[\left(\frac{1}{r_0}\right)^2\right] \ .
\eeq

Once again we can refer to the results of Section \ref{conv} to  estimate the rate of convergence of our series. In this case
the potential is 
\beq
V(z) = z^2 - \frac{2 GM}{r_0} z^3 + \frac{Q^2}{r_0^2} z^4 \ ,
\eeq
and 
\beq
\Delta_{max} =  \frac{GM/r_0-Q^2/r_0^2}{2 - 5 GM/r_0 +3 Q^2/r_0^2} \ .
\eeq

In this case,  the condition of applicability of our series, $\lambda_{max} > \lambda_C$, can
be fulfilled only for $r_0 > 3 GM/2 \left(1 + \sqrt{1-8/9 (Q/GM)^2}\right)$, which is the exact location of the photon sphere for the 
Reissner-Nordstr\"om metric (see eq.(8) of \cite{Eir02}).

\subsection{Janis-Newman-Winicour metric}

We consider now the spherically symmetric metric solution to the Einstein massless scalar equations~\cite{JNW68}:
\beq
A(r) = \left(1-b/r\right)^{-\nu} \ \ \ , \ \ \ B(r) = \left(1-b/r\right)^{\nu} \ \ \ , \ \ \ 
D(r) = \left(1-b/r\right)^{1-\nu} \ ,
\eeq
which reduces to the Schwarzschid metric for $\nu =1$ and for $b = GM$.
In this case we obtain the potential
\beq
V(z) &=& -\left(1-\frac{b}{r_0}\right)^{2 \nu -1} \left(1-\frac{b z}{r_0}\right)^{2-2 \nu }+z^2 
\left(1-\frac{b z}{r_0}\right)+\left(1-\frac{b}{r_0}\right)^{2 \nu -1}
\eeq
which can be expanded around $z=0$ to give
\beq
V(z) &\approx& -2 (\nu -1) (1-\frac{b}{r_0})^{2 \nu -1} \frac{b}{r_0} z + \left[1-(\nu -1) (2 \nu -1) \left(1-\frac{b}{r_0}\right)^{2 \nu -1} 
\left(\frac{b}{r_0}\right)^2\right] \ z^2  \nonumber \\
&+& \left[-\frac{2}{3} (\nu -1) \nu  (2 \nu -1) \left(\frac{b}{r_0}\right)^3 \left(1-\frac{b}{r_0}\right)^{2 \nu -1}-\frac{b}{r_0}\right] \ z^3 
+ O\left[z^4\right] \ .
\eeq

Using this expansion inside our formula, eq.~(\ref{eq_1_17}), we obtain the deflection angle
\beq
\Delta\phi_{PMS}^{(1)} = \pi \ \left[ \frac{1}{\sqrt{1+ \frac{(\nu -1) (1-\xi )^{2 \nu } \xi  ((2 \nu -1) \xi  (8 \nu  \xi +3 \pi \
)+12)-12 (\xi -1) \xi }{3 \pi  (\xi -1)}}} 1-\right]
\label{eq_JNW}
\eeq
where $\xi \equiv b/r_0$.

\begin{figure}
\begin{center}
\includegraphics[width=9cm]{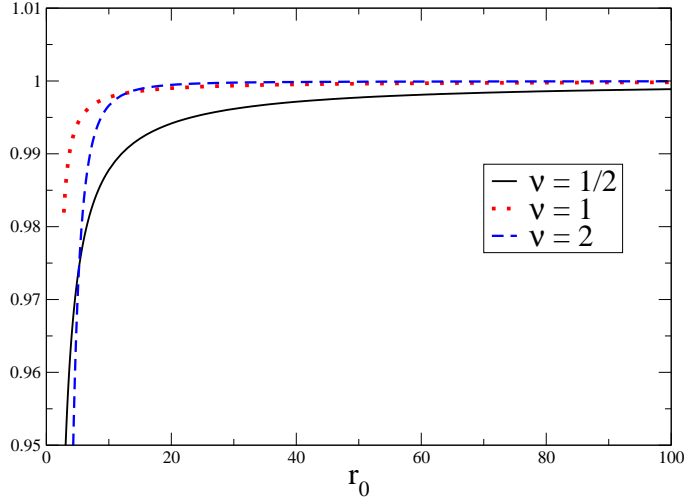}
\bigskip
\caption{Ratio between the approximate first order formula and the exact (numerical) result in the JNW metric 
for different values of $\nu$ and assuming $b=1$. The potential is expanded to order $z^3$. (color online)}
\label{FIGJNW}
\bigskip
\end{center}
\end{figure}

In fig.~\ref{FIGJNW} we have plotted the ratio $\Delta\phi_{PMS}^{(1)}/\Delta\phi_{exact}$ for three values of $\nu$, up to 
small values of $r_0$. Notice that the exact result is calculated numerically and that the ratio is close to $1$ up to very 
small values of $r_0$.

The expansion of eq.~(\ref{eq_JNW}) around $r_0$ reads
\beq
\Delta\phi_{PMS}^{(1)} \approx \frac{2b \nu}{r_0} + 
\frac{\left(2 \left(6-4 \pi +\pi ^2\right) \nu ^2-3 (-4+\pi ) \pi  \nu +(-4+\pi ) \pi \right)}{2 \pi } \frac{b^2}{r_0^2} + 
O\left[1/r_0^3\right] \ ,
\eeq
and provides a deviation from Einstein's leading order term, $\Delta\phi \approx 2b/r_0$ for $\nu \neq 1$.

\subsection{Einstein-Born-Infeld black holes}

As a final example of application of our method we consider the propagation of light in a charged black 
hole coupled to Born-Infeld electrodynamics. This problem has been recently considered by Eiroa in 
\cite{Eir06} and corresponds to the effective metric 
\beq
A(r) = \frac{\sqrt{\omega(r)}}{\psi(r)} \ \ , \ \ B(r) = \sqrt{\omega(r)} \psi(r) \ \ , \ \  D(r) = 
\frac{1}{\sqrt{\omega(r)}} \ ,
\eeq
where 
\beq
\omega(r) &=& 1 + \frac{Q^2 b^2}{r^4} \\
\psi(r)   &=& 1- 2 \frac{M}{r} + \frac{2}{3 b^2} \ \left\{ r^2 - \sqrt{r^4+b^2 Q^2} + \frac{\sqrt{|bQ|^3}}{r} \ 
F\left[\arccos \left( \frac{r^2-|bQ|}{r^2+|bQ|} \right) , \frac{1}{\sqrt{2}} \right] \right\}  \ .
\eeq
$F(a,b)$ is the incomplete elliptic integral of first kind. We follow the convention of \cite{Eir06} and set $G=1$.

In this case one obtains the potential
\beq
V(z) &\equiv& z^2  \frac{\psi(r_0/z)}{ \omega(r_0/z)} - \frac{\psi(r_0) \omega(r_0)}{\omega^2(r_0/z)} + 
\psi(r_0) \omega(r_0)
\eeq
which can be expanded around $z=0$ as
\beq
V(z) = \sum_{n=2}^\infty v_n z^n \ .
\label{eq_series}
\eeq

Unlike in the previous two cases $V(z)$ {\sl is not polynomial } in $z$ and the deflection angle 
reads
\beq
\Delta\phi_{PMS}^{(1)} &=&  \pi \ \left[ \frac{1}{\sqrt{1- \frac{8 M}{\pi r_0} + \frac{3}{2} v_4 + 
\frac{16}{3\pi} v_5 + \frac{15}{8} v_6 + \dots }} -1
\right]
\eeq
Clearly one has to keep in mind that the truncation of the series (\ref{eq_series}) to a finite order 
is an additional source of error in our calculation: in practice, however, it is straightforward to 
include further terms of the expansion.

\begin{figure}
\begin{center}
\includegraphics[width=9cm]{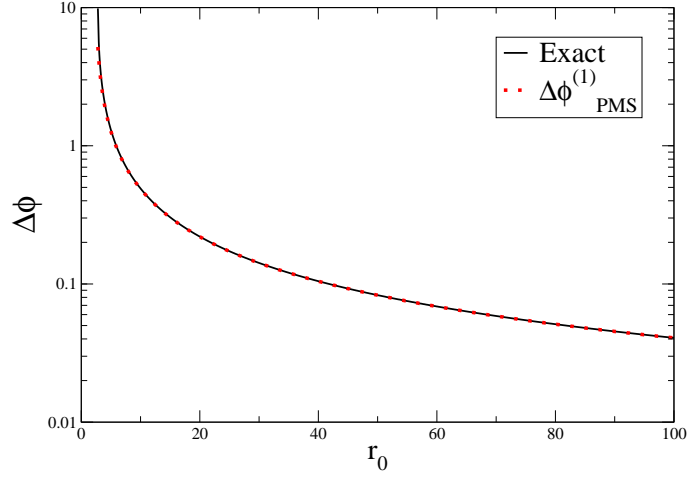}
\bigskip
\caption{Deflection angle for the Einstein-Born-Infeld problem as a function of $r_0$ assuming 
$b=M=1$ and $Q=1/2$. The potential in our formula is expanded to order $z^4$. (color online)}
\label{FIG4}
\bigskip
\end{center}
\end{figure}

In Fig.~\ref{FIG4} we have compared the exact result obtained {\sl numerically} integrating 
the integral in $\Delta\phi$ with the result obtained with our first order formula using  
the expansion $V(z)$ to order $z^4$:
\beq
V(z) \approx z^2 - \frac{2 M}{r_0} z^3  + v_4 z^4 + O[z^5]
\eeq
where
\beq
v_4 &=& \frac{7 Q^2}{3 r_0^2}+ \frac{2}{3} \left(3 b^2-2 \beta \right) \frac{Q^2}{r_0^4} + 
\frac{4}{3} \frac{b Q^2}{r_0^5} \left(\gamma  Q \sqrt{b Q}-3 b M\right) + \frac{4 b^2 Q^4}{3 r_0^6} \nonumber \\
&+& \frac{2}{3} b^2 \left(3 b^2-2 \beta \right) \frac{Q^4}{r_0^8} + 
\frac{4 M b^4 Q^4}{r_0^9} \ \left(\frac{\gamma  Q^{3/2}}{3 \sqrt{b} M}-1
\right) 
\eeq
and
\beq
\beta^2 &\equiv& b^2Q^2+r_0^4 \\
\gamma &\equiv& F\left(\arccos\left(1-\frac{2 b Q}{r_0^2+b Q}\right),\frac{1}{\sqrt{2}}\right) \ .
\eeq
Our {\sl analytical formula} reproduces with high accuracy the numerical result obtained assuming
$b=M=1$ and $Q=1/2$.

It is also easy to obtain the asymptotic behavior of $\Delta\phi$ from our expression
\beq
\Delta\phi \approx \frac{4M}{r_0} + \left(\frac{24}{\pi} \frac{M^2}{r_0^2} - \frac{3\pi}{4} 
\frac{Q^2}{r_0^2}\right) + \left( \frac{160}{\pi^2} \frac{M^3}{r_0^3} - 9 \frac{M Q^2}{r_0^3}  
\right) + O \left[\frac{1}{r_0^4}\right] \ .
\eeq

\section{Conclusions}
\label{concl}

In this paper we have presented a new method to obtain analytical expressions
for the deflection angle of a ray of light propagating in a spherically symmetric 
static metric. We have been able to prove the convergence of our approach and
to estimate the rate of convergence of the series obtained applying our
method: the series converges exponentially and can be applied over all the 
physical region, as explictly seen in the case of the Schwarzschild and 
Reissner-Nordstr\"om metrics, where the correct location of the photon sphere
is recovered.

This method has been used to derive a first order formula, which is valid for 
general spherically symmetric static metric tensor: we have tested this formula 
in four different cases,  observing that it is quite accurate even in proximity 
of the photon sphere. Clearly, higher order corrections to the first order formula 
of this paper will further improve the quality approximation, given the convergent 
nature of our expansion: we plan to study higher order corrections to our formula 
in a forthcoming paper.
 
We also stress that the series obtained with our method are nonperturbative, because 
they do not correspond  to an expansion in a small parameter and therefore they are 
capable of providing small errors even when the parameters in the model are not small 
(a typical perturbative parameter would be  $GM/r_0$).

\begin{acknowledgments}
P.A. acknowledges support of Conacyt grant no. C01-40633/A-1.
\end{acknowledgments}


\begin{thebibliography}{}
\bibitem{FKN00} S. Frittelli, T.P. Kling and T.Newman, Phys.Rev.{\bf D} 61, 064021 (2000)
\bibitem{VE00} K.S.Virbhadra and G.F.R.Ellis,Phys. Rev. {\bf D} 62, 084003 (2000)
\bibitem{VE02} K.S.Virbhadra and G.F.R.Ellis,Phys. Rev. {\bf D} 65, 103004 (2002)
\bibitem{Eir02} E.F.Eiroa,G.E.Romero and D.F.Torres, Phys. Rev. {\bf D} 66, 024010 (2002)
\bibitem{Bha03} A. Bhadra, Phys. Rev. {\bf D} 67, 103009 (2003)
\bibitem{Boz03} V. Bozza, Phys. Rev. {\bf D} 67, 103006 (2003);
                V. Bozza, F. De Luca, G. Scarpetta, M. Sereno, Phys.Rev. {\bf D}72, 08300 (2005)
\bibitem{Whi05} R. Whisker, Phys. Rev. {\bf D} 71, 064004 (2005)
\bibitem{Eir05} E.F. Eiroa, Phys. Rev. {\bf D} 71, 083010 (2005)
\bibitem{Eir06} E.F.Eiroa, Phys. Rev. {\bf D} 72, 043002 (2006)
\bibitem{SB06} K. Sarkar and A. Bhadra, ArXiv:[gr-qc/0602087] (2006)
\bibitem{Perl04} V. Perlick, Phys.Rev. {\bf D} 69, 064017 (2004)
\bibitem{Boz02} V.Bozza, Phys. Rev. {\bf D} 66, 103001 (2002)
\bibitem{Mutkaa} P.T. Mutka and P. M\"ah\"onen, The Astrophysical Journal {\bf 581}: 1328-1336 (2002)
\bibitem{Mutkab} P.T. Mutka and P. M\"ah\"onen, The Astrophysical Journal {\bf 576}: 107-112 (2002)
\bibitem{Belo02} A.M. Beloborodov, The Astrophysical Journal {\bf 566}: L85-L88 (2002)
\bibitem{Keet05} C.R. Keeton and A.O. Petters, Phys. Rev. {\bf D} 72, 104006 (2005)
\bibitem{Am05a} P.Amore and R.A.Sa\'enz, Europhysics letters {\bf 70} 425-431 (2005)
\bibitem{Am05b} P.Amore, A.Aranda, F.Fernandez and R.A.Sa\'enz, Phys. Rev.{\bf E} 71 (2005)
\bibitem{lde}   A. Okopi\'nska, Phys.\ Rev.\ D {\bf 35}, 1835 (1987); 
                A. Duncan and M. Moshe, Phys.\ Lett.\ B {\bf 215}, 352 (1988);
                H.F.Jones and M. Moshe, Phys.\ Lett.\ B {\bf 234}, 492 (1990)
\bibitem{VPT} H. Kleinert, Path Integrals in Quantum Mechanics, 
                 Statistics and Polymer Physics, 3rd edition (World Scientific Publishing, 2004)
\bibitem{Ste81} P.M. Stevenson, Phys. Rev. D {\bf 23}, 2916 (1981)
\bibitem{Weinberg}       S. Weinberg, Gravitation and cosmology, J.Wiley and Sons, 1972
\bibitem{Darw59} C. Darwin, Proc.R. Soc. London {\bf A} 249, 180 (1959); C. Darwin, Proc.R. Soc. London {\bf A} 263, 39 (1961);
\bibitem{ADL06} P. Amore, A. De Pace and J. Lopez, submitted to Jour. of Phys. {\bf G}: ArXiv:[hep-ph/0602114]  (2006)
\bibitem{AmJMAA} P. Amore, accepted on the Journal of Mathematical analysis and applications, ArXiv:[math-ph/0408036], (2006)
\bibitem{AmJPA} P. Amore, J. of Phys. {\bf A} 38, 6463-6472 (2005) 




\bibitem{JNW68} A.I.Janis, E.T.Newman and J.Winicour, Phys.Rev.Lett. {\bf 20}, 878 (1968)

\end{thebibliography}
\end{document}